\documentclass[11pt,a4paper]{amsart}
\usepackage[margin=25mm]{geometry}
\usepackage[numbers]{natbib}
\usepackage{hyperref}

\usepackage{amssymb,amsmath}
\usepackage{amsthm}
\usepackage{bbm}
\usepackage{stmaryrd}
\usepackage[usenames,dvipsnames]{xcolor}
\usepackage{color}
\usepackage{graphicx}
\usepackage{caption}
\usepackage{float}

\usepackage{lineno}

\usepackage{abstract}

\input xy
\xyoption{all} 

\numberwithin{equation}{section}

\theoremstyle{definition}

\newenvironment{warning}[1][Warning.]{\begin{trivlist}
\item[\hskip \labelsep {\bfseries #1}]}{\end{trivlist}}

\newenvironment{remark}[1][Remark.]{\begin{trivlist}
\item[\hskip \labelsep {\bfseries #1}]  }{ \end{trivlist}}

\newcommand{\rme}{\textnormal{e}}

\font\black=cmbx10 \font\sblack=cmbx7 \font\ssblack=cmbx5 \font\blackital=cmmib10  \skewchar\blackital='177
\font\sblackital=cmmib7 \skewchar\sblackital='177 \font\ssblackital=cmmib5 \skewchar\ssblackital='177
\font\sanss=cmss10 \font\ssanss=cmss8 
\font\sssanss=cmss8 scaled 600 \font\blackboard=msbm10 \font\sblackboard=msbm7 \font\ssblackboard=msbm5
\font\caligr=eusm10 \font\scaligr=eusm7 \font\sscaligr=eusm5  \font\fraktur=eufm10
\font\sfraktur=eufm7 \font\ssfraktur=eufm5 
\font\bsymb=cmsy10 scaled\magstep2
\def\all#1{\setbox0=\hbox{\lower1.5pt\hbox{\bsymb
       \char"38}}\setbox1=\hbox{$_{#1}$} \box0\lower2pt\box1\;}
\def\exi#1{\setbox0=\hbox{\lower1.5pt\hbox{\bsymb \char"39}}
       \setbox1=\hbox{$_{#1}$} \box0\lower2pt\box1\;}

\def\tx#1{{\fam0\relax#1}}

\newfam\bifam
\textfont\bifam=\blackital \scriptfont\bifam=\sblackital \scriptscriptfont\bifam=\ssblackital

\newfam\blfam
\textfont\blfam=\black \scriptfont\blfam=\sblack \scriptscriptfont\blfam=\ssblack

\newfam\bbfam
\textfont\bbfam=\blackboard \scriptfont\bbfam=\sblackboard \scriptscriptfont\bbfam=\ssblackboard

\newfam\ssfam
\textfont\ssfam=\sanss \scriptfont\ssfam=\ssanss \scriptscriptfont\ssfam=\sssanss

\newfam\clfam
\textfont\clfam=\caligr \scriptfont\clfam=\scaligr \scriptscriptfont\clfam=\sscaligr

\newfam\frfam
\textfont\frfam=\fraktur \scriptfont\frfam=\sfraktur \scriptscriptfont\frfam=\ssfraktur

\def\hpb#1{\setbox0=\hbox{${#1}$}
    \copy0 \kern-\wd0 \kern.2pt \box0}
\def\vpb#1{\setbox0=\hbox{${#1}$}
    \copy0 \kern-\wd0 \raise.08pt \box0}

\def\pmb#1{\setbox0\hbox{${#1}$} \copy0 \kern-\wd0 \kern.2pt \box0}
\def\pmbb#1{\setbox0\hbox{${#1}$} \copy0 \kern-\wd0
      \kern.2pt \copy0 \kern-\wd0 \kern.2pt \box0}
\def\pmbbb#1{\setbox0\hbox{${#1}$} \copy0 \kern-\wd0
      \kern.2pt \copy0 \kern-\wd0 \kern.2pt
    \copy0 \kern-\wd0 \kern.2pt \box0}
\def\pmxb#1{\setbox0\hbox{${#1}$} \copy0 \kern-\wd0
      \kern.2pt \copy0 \kern-\wd0 \kern.2pt
      \copy0 \kern-\wd0 \kern.2pt \copy0 \kern-\wd0 \kern.2pt \box0}
\def\pmxbb#1{\setbox0\hbox{${#1}$} \copy0 \kern-\wd0 \kern.2pt
      \copy0 \kern-\wd0 \kern.2pt
      \copy0 \kern-\wd0 \kern.2pt \copy0 \kern-\wd0 \kern.2pt
      \copy0 \kern-\wd0 \kern.2pt \box0}

\mathchardef\za="710B  
\mathchardef\zb="710C  
\mathchardef\zg="710D  
\mathchardef\zd="710E  
\mathchardef\zve="710F 
\mathchardef\zz="7110  
\mathchardef\zh="7111  
\mathchardef\zvy="7112 
\mathchardef\zi="7113  
\mathchardef\zk="7114  
\mathchardef\zl="7115  
\mathchardef\zm="7116  
\mathchardef\zn="7117  
\mathchardef\zx="7118  
\mathchardef\zp="7119  
\mathchardef\zr="711A  
\mathchardef\zs="711B  
\mathchardef\zt="711C  
\mathchardef\zu="711D  
\mathchardef\zvf="711E 
\mathchardef\zq="711F  
\mathchardef\zc="7120  
\mathchardef\zw="7121  
\mathchardef\ze="7122  
\mathchardef\zy="7123  
\mathchardef\zf="7124  
\mathchardef\zvr="7125 
\mathchardef\zvs="7126 
\mathchardef\zf="7127  
\mathchardef\zG="7000  
\mathchardef\zD="7001  
\mathchardef\zY="7002  
\mathchardef\zL="7003  
\mathchardef\zX="7004  
\mathchardef\zP="7005  
\mathchardef\zS="7006  
\mathchardef\zU="7007  
\mathchardef\zF="7008  
\mathchardef\zW="700A  
\mathchardef\zC="7009  

\newcommand{\be}{\begin{equation}}
\newcommand{\ee}{\end{equation}}

\newcommand{\bea}{\begin{eqnarray}}
\newcommand{\eea}{\end{eqnarray}}
\def\*{{\textstyle *}}
\newcommand{\R}{{\mathbb R}}

\newcommand{\Z}{{\mathbb Z}}

\newcommand{\s}{{\textstyle *}}

\def\xi{\tx{i}}

\def\s*{{\scriptstyle *}}

\def\cO{\mathcal{O}}

\newcommand{\beas}{\begin{eqnarray*}}
\newcommand{\eeas}{\end{eqnarray*}}

\def\half{\frac{1}{2}}

\title{Is the $\Z_2\times \Z_2$-graded sine-Gordon equation integrable?} 
\author{Andrew James Bruce } 
   \address{Mathematics Research Unit, University of Luxembourg, Maison du Nombre 6, avenue de la Fonte, 
L-4364 Esch-sur-Alzette}  
   \email{andrewjamesbruce@googlemail.com}

   \date{\today}
 
\begin{document}
 \maketitle
\vspace{-20pt}
\begin{abstract}{\noindent We examine the question of the integrability of the recently defined $\Z_2\times \Z_2$-graded sine-Gordon model, which is a natural generalisation of the supersymmetric sine-Gordon equation. We do this via appropriate auto-B\"{a}cklund transformations, construction of conserved spinor-valued currents and  a pair of infinite sets of conservation laws. }\\

\noindent {\Small \textbf{Keywords:} Integrable Systems;~ Supersymmetric Models;~B\"{a}cklund Transformations;~$\Z_2 \times \Z_2$-Grading}
\end{abstract}

\section{Introduction} 
There has been some renewed interest in $\Z_2\times \Z_2$-gradings (and higher) in physics, see for example \cite{Aizawa:2020a,Aizawa:2020,Aizawa:2021,Bruce:2020a,Toppan:2021} and references therein. It has long been appreciated that these ``higher gradings'' play a r\^ole in the theory of parastatistics, relatively recent works in this area include Tolstoy \cite{Tolstoy:2014} (also see references therein). In particular, Toppan in \cite{Toppan:2021} has shown that in multi-particle theories it is possible to distinguish between ordinary bosons/fermions, so $\Z_2$-graded particles, and the exotic $\Z_2 \times \Z_2$-graded particles. It is thus conceivable that there is  new physics behind these novel theories, especially in low-dimensional condensed matter physics where the spin-statistics theorem does not hold. In this paper, we continue the study of a $1+1$-dimensional classical field theory that is $\Z_2\times \Z_2$-graded.\par
Sigma models that exhibit a novel $\Z_2\times \Z_2$-graded generalisation of supersymmetry (see \cite{Bruce:2019a}) were first presented by the author in \cite{Bruce:2020}. As an example of one of these sigma models a $\Z_2 \times \Z_2$-graded version of the supersymmetric sine-Gordon theory was given and some initial study was made. In this short paper, we examine the natural question of the integrability of the  $\Z_2 \times \Z_2$-graded sine-Gordon equation.  There is no universal definition of integrability of systems with an infinite number of degrees of freedom. We will use the definition that a classical field theory is (formally) integrable if there is an infinite set of conserved currents (equivalently, conservation laws).  Our approach to the question of integrability is to construct  relevant auto-B\"{a}cklund transformations and from there build the infinite set of conservation laws. The standard supersymmetric case was examined by Chaichian \& Kulish \cite{Chainchain:1978} using a superspace auto-B\"{a}cklund transformation following earlier calculations of the infinite set of conserved currents by  Ferrara, Girardello \&  Sciuto \cite{Ferrara:1978}. There are some complications with the  $\Z_2\times \Z_2$-graded case due to the presence of a degree $(1,1)$ parameter in the Lagrangian and the subsequent need for spinor-valued parameters of degree $(0,1)$ and $(1,0)$. There is no general systematic method for constructing B\"acklund transformations and as such we take our cue from the classical transformations for the standard sine-Gordon equation making the necessary modifications.  \par 
Other approaches to the integrability of supersymmetric sine-Gordon exist, such as constructing Lax pairs via superconformally affine Toda theories (see \cite{Evans:1992,Ivanov:1993}) and the construction of solitons (see for example \cite{DiVecchia:1977,Liu:1998}). For a study of the different aspects of the integrability of the supersymmetric sine-Gordon see Bertrand \cite{Bertrand:2017}. It would be interesting to see how these other approaches generalise to this ``higher graded'' setting. \par 
We remark that the sine-Gordon model is found throughout mathematical physics and has a very wide range of applications such as coupled rigid pendula, Josephson junctions, crystal dislocations and 2-d gravitational physics. A review of some of these applications can be found in \cite{Barone:1971}. Furthermore, $\mathcal N= 2$ supersymmetric sine–Gordon models can be found in string theory via  reductions of superstring worldsheet theories on particular backgrounds, see for example \cite{Grigoriev:2008}. More generally, two-dimensional field theories appear in condensed matter physics and statistical mechanics while often exhibiting  the mathematical  features of four-dimensional theories.

\section{$\Z_2\times \Z_2$-supersymmetry}
The \emph{$\Z_2 \times \Z_2$-graded, $d=2$, $\mathcal{N}= (1,1)$ supertranslation algebra} (in light-cone coordinates)  is the $\Z_2\times \Z_2$-Lie superalgebra (see \cite{Rittenberg:1978,Scheunert:1979}) with $5$ generators of  $\Z_2 \times \Z_2$-degree
$$\{\underbrace{P_-}_{(0,0)}, ~ \underbrace{P_+}_{(0,0)}, ~ \underbrace{Z_{-+}}_{(1,1)},~\underbrace{Q_-}_{(0,1)} ,~ \underbrace{Q_+}_{(1,0)} \} ,$$
where they transform under $2$-d Lorentz boosts (here $\beta \in \R$ is the rapidity) as
\begin{align*}
& P_-   \mapsto \rme^\beta \, P_-, 
& P_+   \mapsto \rme^{-\beta} \, P_+,
&& Z_{-+} \mapsto Z_{-+}. &\\
& Q_- \mapsto \rme^{\half \beta}\, Q_-,
& Q_+ \mapsto \rme^{-\half \beta}\, Q_+.&&\\
\end{align*}  
\begin{warning}
Thinking of the following $\Z_2 \times \Z_2$-Lie superalgebra \eqref{eqn:ZZLieAlg} as being embedded in an enveloping algebra, the first two brackets are, in the classical language, anticommutators, while the third is a commutator. The nature of the brackets is determined by the $\Z_2\times \Z_2$-degree of the elements and the map $\Z_2\times \Z_2 \ni (\mathbf{a}, \mathbf{b})\mapsto \langle\mathbf{a}, \mathbf{b}  \rangle = a_1 \,b_1 + a_2 \, b_2 \, \mod 2 \in \{ 0,1\}$. Supposing we have two elements  $A$ and $B$ from some enveloping algebra of $\Z_2\times \Z_2$-degree $\mathbf{a}$ and $\mathbf{b}$, respectively, the graded Lie bracket is $[A,B]= A\cdot B -(-1)^{\langle \mathbf{a}, \mathbf{b}\rangle}\, B \cdot A$. The graded Jacobi identity is $[A, [B,C]] = [[A, B], C]+ (-1)^{\langle \mathbf{a}, \mathbf{b}\rangle}\, [B, [A,C]]$.  Note that we have the natural notion of the total degree of an element but this is not sufficient to determine the required sign factors. For example, given a pair of odd elements with $\Z_2\times\Z_2$-degree $(0,1)$ and $(1,0)$, respectively, it is the commutator and not the anticommutator that we are interested in.   This is, of course, quite different to standard supersymmetry. For an overview of supersymmetry and supermathematics the reader can consult \cite{Duplij:2004}.
\end{warning}
The  non-vanishing $\Z_2 \times \Z_2$-graded Lie brackets are
\begin{align}\label{eqn:ZZLieAlg}
& [Q_-, Q_-] = P_-, && [Q_+, Q_+] = P_+, & [Q_-, Q_+] = Z_{-+}.
\end{align}
We see that the generators consist of a pair of right-handed and left-handed vectors, a  pair of right-handed and left-handed  Majorana--Weyl spinors and a Lorentz scalar.\par 
The \emph{$d=2$, $\mathcal{N} = (1,1)$ $\Z_2 \times \Z_2$-Poincar\'e algebra} is the above $\Z_2 \times \Z_2$-Lie superalgebra with an additional degree $(0,0)$ generator $M$, understood as the generator of Lorentz boosts, together with the non-vanishing   $\Z_2 \times \Z_2$-graded Lie brackets (all other brackets are zero)
\begin{equation}
[M, P_\mp] =  \mp P_{\pm}\, , \quad [M, Q_\pm] =  \mp \frac{1}{2} Q_\pm \,.
\end{equation}
The geometric realisation of this novel supersymmetry is via \emph{$d=2$, $\mathcal{N} = (1,1)$ $\Z_2 \times \Z_2$-Minkowski spacetime}, which is the $\Z_2 \times \Z_2$-manifold (see \cite{Covolo:2016})  that comes equipped with global coordinate systems of the form
 $$\{\underbrace{x^-}_{(0,0)}, ~ \underbrace{x^+}_{(0,0)},  ~ \underbrace{z}_{(1,1)},~\underbrace{\theta_-}_{(0,1)} ,~ \underbrace{\theta_+}_{(1,0)}  \}\,.$$
Under Lorentz boosts these coordinate transform as 
\begin{align*}
& x^-   \mapsto \rme^{-\beta} \, x^-, 
& x^+   \mapsto \rme^{\beta} \, x^+,
&& z \mapsto z, &\\
& \theta_- \mapsto \rme^{-\half \beta}\, \theta_-,
& \theta_+ \mapsto \rme^{\half \beta}\, \theta_+,&&
\end{align*}  
again, $\beta$ is the rapidity. Thus, the underlying reduced manifold is two-dimensional Minkowski spacetime, here explicitly given in light-cone coordinates. \par  
The following vector fields form a representation of the $\Z_2 \times \Z_2$-supertranslation algebra 
\begin{align}
 \nonumber & P_- = \frac{\partial}{\partial x^-}, \quad P_+ = \frac{\partial}{\partial x^+},\quad Z_{-+} = \frac{\partial}{\partial z},\\
& Q_- = \frac{\partial }{\partial \theta_-} + \frac{1}{2} \theta_- \frac{\partial}{\partial x^-}  - \frac{1}{2} \theta_+ \frac{\partial}{\partial z},\quad
 Q_+ = \frac{\partial }{\partial \theta_+}  + \frac{1}{2} \theta_+ \frac{\partial}{\partial x^+} + \frac{1}{2} \theta_- \frac{\partial}{\partial z}.
\end{align}
The reader can easily verify that these vector fields have the right transformation properties under a Lorentz boost. \par
In standard supersymmetry, the partial derivatives for the odd coordinates of superspace are not invariant under supertranslations and so we need covariant derivatives. The required $\Z_2 \times \Z_2$-SUSY \emph{covariant derivatives} are
\begin{align}
& D_- = \frac{\partial }{\partial \theta_-} - \frac{1}{2} \theta_- \frac{\partial}{\partial x^-}  +  \frac{1}{2}\theta_+  \frac{\partial}{\partial z},
& D_+ = \frac{\partial }{\partial \theta_+} - \frac{1}{2} \theta_+ \frac{\partial}{\partial x^+} - \frac{1}{2} \theta_- \frac{\partial}{\partial z}.
\end{align}
Direct calculation gives the expected results
$$ [D_\mp, D_\mp] = - P_\mp \,, \quad  [D_-, D_+] = -Z_{-+}\,. $$
We will take  $\Z_2 \times \Z_2$-superfields to be scalar under the $2$-d Lorentz transformations and of $\Z_2 \times \Z_2$-degree zero. Then to first order in $z$,
\begin{align}
 \nonumber & \Phi(x^-,x^+ ,z,\theta_- , \theta_+) = X(x^-,x^+) + \theta_- \, \psi_+(x^-,x^+)\\ 
\nonumber  & + \theta_+ \, \psi_-(x^-,x^+) + \theta_- \theta_+ \ F_{+-}(x^-,x^+)\\ 
\nonumber &+ z G(x^-,x^+) + z \theta_- \,\chi_+(x^-,x^+) + z \theta_+ \,\chi_-(x^-,x^+)\\
&+  z \theta_- \theta_+\,  Y_{+-}(x^-,x^+) \, + \mathcal{O}(z^2).
\end{align}
In general, we have a formal power series in $z$, and so we have an infinite number of component fields. The first four types of component  fields  are
$$\big( \underbrace{X}_{(0,0)},~  \underbrace{F_{+-}}_{(1,1)},~ \underbrace{\psi_+}_{(0,1)}, ~  \underbrace{\psi_-}_{(1,0)}\big). $$
For the $\Z_2 \times \Z_2$-superfield to be a Lorentz scalar we require that, under Lorentz boosts
\begin{align*}
& \psi_+ \mapsto \rme^{\half \beta}\,\psi_+ , && \psi_- \mapsto \rme^{-\half \beta}\,\psi_-,
\end{align*} 
while $X$ and $F_{+-}$ are Lorentz scalars. That is we have  bosons, exotic boson that anticommute with the fermions while commuting with the bosons, and  left-handed and right-handed  fermions that mutually commute. This is quite different to the component fields on standard superspace. For more details the reader should consult \cite{Bruce:2019a,Bruce:2020}.
\begin{remark}
We will be mathematically lax in our definition of a $\Z_2\times \Z_2$-superfield and simply allow the components thereof to depend on extra unspecified  $\Z_2\times \Z_2$-graded parameters as required. Furthermore, it is possible to consider $\Z_2\times \Z_2$-superfield of non-zero $\Z_2\times \Z_2$-degree and of a different nature with respect to Lorentz transformations.  
\end{remark}
\section{The $\Z_2\times \Z_2$-graded sine-Gordon equation}
The $\Z_2 \times \Z_2$-space Euler--Lagrange equations\footnote{At this point we postulate these equations.} are
\begin{equation}
D_-\left( \frac{\partial L}{\partial \Phi_-}\right) +D_+\left( \frac{\partial L}{\partial\Phi_+}\right)- \frac{\partial W}{\partial \Phi} = 0\,,
\end{equation}
where $\Phi_{\mp} = D_{\mp}\Phi$ and $L := L[\Phi, \Phi_-, \Phi_+]$ is a  scalar functional, we will comment on the $\Z_2 \times \Z_2$-degree shortly. Here $W :=  W[\Phi]$ is a scalar potential. The $\Z_2 \times \Z_2$-graded sine-Gordon Lagrangian is defined as
\begin{equation}
L[\Phi, \Phi_-, \Phi_+] = D_- \Phi D_+ \Phi - 2 \alpha \, \big(1- \cos(\Phi/2)  \big)\,,
\end{equation}
where $\alpha$ is a $\Z_2 \times \Z_2$-degree $(1,1)$ constant that satisfies $\alpha^2 - 1 =0$. This constant was introduced because the Lagrangian should be homogeneous and of $\Z_2\times \Z_2$-degree $(1,1)$, this implies that the action is $\Z_2\times \Z_2$-degree $(0,0)$. The Clifford-relation is imposed so that this constant does not appear in the underlying classical equations of motion, in this case the classical sine-Gordon equation (see \cite{Bruce:2020} for further details).
\begin{remark}
Coupling constants that are of $\Z_2 \times \Z_2$-degree $(1,1)$ appear in $\Z_2 \times \Z_2$-graded classical mechanics for exactly the same reason they appear in $2$-dimensional field theories (see \cite{Aizawa:2020}). It is thus expected that non-trivial dynamics of $\Z_2^n$-graded classical theories requires non-zero graded coupling constants. In the context of finding non-supersymmetric finite field theories, Clifford algebra-valued coupling constants were first conjectured by Kranner \& Kummer (see \cite{Kranner:1991}). Supergroups with Clifford algebra-valued parameters together with their relation with non-anticommuting supersymmetry were discussed by Kuznetsova, Rojas, \& Toppan (see \cite{Kuznetsova:2008}).
\end{remark}
 The \emph{$\Z_2 \times \Z_2$-graded sine-Gordon equation} is thus
\begin{equation}
D_- D_+ \Phi + D_+ D_- \Phi + \alpha \, \sin\left(\frac{\Phi}{2} \right) =0.
\end{equation}
An initial analysis of the $\Z_2 \times \Z_2$-graded sine-Gordon system was given in \cite{Bruce:2020} including carefully showing how it is invariant under $\Z_2 \times \Z_2$-graded supersymmetry (\cite{Bruce:2019a}) and the explicit presentation of the associated current.
\begin{remark}
For the integration theory to be sound there are constraints on the $\Z_2 \times \Z_2$-superfield.  It should be noted that the integration theory on $\Z_2^n$-manifolds is not well-developed. As we will not directly use the integration theory we will not discuss this further.  For details see \cite{Bruce:2020,Poncin:2016}. 
\end{remark}
The equations of motion for the lowest-order component fields (eliminating the auxiliary field using its equation of motion  $2 F_{+-}= -  \alpha\, \sin(X/2)$ )  are 
\begin{align}\label{Eqn:Z22SinGordEqCom}
 \nonumber &\partial_- \partial_+ X - \frac{1}{4}  \sin(X) - \frac{\alpha}{2} \,\psi_- \psi_+ \, \sin(X/2)=0,&&\\
& \partial_+ \psi_+  + \frac{\alpha}{2} \, \psi_- \cos(X/2)=0,
&& \partial_- \psi_- +  \frac{\alpha}{2} \, \psi_+ \cos(X/2)=0.
 \end{align}
Setting $\psi_+ = \psi_- =0$ reduces the system to the classical sine-Gordon equation. Thus, the classical solutions, for example, the topological kinks and anti-kinks are solutions to the $\Z_2 \times \Z_2$-graded sine-Gordon equation once the non-zero degree component fields are set to zero.

\section{ $\Z_2 \times \Z_2$-graded auto-B\"{a}cklund transformations}
An auto-B\"{a}cklund transformation for the $\Z_2\times \Z_2$-supersymmetric sine-Gordon equation is defined as follows: 
\begin{subequations}
\begin{align}
& D_-\widetilde{\Phi} =  D_- \Phi + 2 a \, \lambda_+ \sin\left( \frac{\widetilde{\Phi} + \Phi}{4}\right)\,, \label{eqn:BTa}\\
&  D_+\widetilde{\Phi} = - D_+ \Phi + 2 a^{-1} \, \lambda_- \sin\left( \frac{\widetilde{\Phi} - \Phi}{4}\right)\,, \label{eqn:BTb}\\
& Z_{-+ }\widetilde{\Phi} =  Z_{-+}\Phi\,, \label{eqn:BTc}
\end{align}
\end{subequations}
where $a \in \R_*$, and $\lambda_+$  \& $\lambda_-$ are  constant spinor parameters of degree $(0,1)$ and $(1,0)$, respectively. More carefully, under Lorentz transformations we have 
$$\lambda_+ \mapsto \rme^{\half \beta}\, \lambda_+\, , \quad \lambda_- \mapsto \rme^{- \half \beta} \, \lambda_-\,.$$
Moreover, these spinor-valued parameters must satisfy 
\begin{equation}
\lambda_+ \lambda_- =  \alpha\, , \quad \lambda_- \lambda_+ + \lambda_+ \lambda_- =0\,. 
\end{equation}
We observe that $ (\lambda_+)^2 (\lambda_-)^2 = -1$. We then set (there is a choice here with a sign) $(\lambda_+)^2 =  v_+$ and $(\lambda_-)^2 =  - v_-$, where $v_+$  and $v_-$ are  $\Z_2\times \Z_2$-degree $(0,0)$ vector-valued parameters that are dual, i.e.,
$$v_+ \mapsto \rme^{\beta} \, v_+\,, \quad v_- \mapsto \rme^{-\beta} \, v_-\,,\quad v_+ v_- = v_+ v_- = 1\,.$$
\begin{remark}
There is another natural auto-B\"{a}cklund upon changing $D_-$ and $D_+$. We will use this other transformation later.
\end{remark}
We claim that if the field $\Phi$ is a solution to the $\Z_2\times \Z_2$-supersymmetric sine-Gordon equation, then the field $\widetilde{\Phi}$ is also a solution. We show this via direct computation. From \eqref{eqn:BTa} \& \eqref{eqn:BTb} we have
\begin{align*}
& D_+ D_- \widetilde{\Phi} = D_+ D_-\Phi - \lambda_- \lambda_+ \, \cos\left( \frac{\widetilde{\Phi}+ \Phi}{4}\right) \sin \left( \frac{\widetilde{\Phi}- \Phi}{4}\right)\, , \\
& D_- D_+ \widetilde{\Phi} = - D_- D_+\Phi + \lambda_- \lambda_+ \, \cos\left( \frac{\widetilde{\Phi} - \Phi}{4}\right) \sin \left( \frac{\widetilde{\Phi}+ \Phi}{4}\right)\, .
\end{align*}
Adding and subtracting the two equations above produces 
\begin{align*}
&D_+ D_- \widetilde{\Phi} + D_- D_+ \widetilde{\Phi} = Z_{-+} \Phi + \lambda_- \lambda_+ \, \sin \left( \frac{\widetilde{\Phi}}{2}\right)\, ,\\
&Z_{-+}\widetilde{\Phi} = D_+ D_- \Phi + D_- D_+ \Phi - \lambda_-\lambda_+\, \sin \left( \frac{\Phi}{2}\right)\,.
\end{align*}
Using \eqref{eqn:BTc} and $\lambda_+ \lambda_- = \alpha$ we obtain
$$ D_+ D_- \widetilde{\Phi} + D_- D_+ \widetilde{\Phi} + \alpha \, \sin\left(\frac{\widetilde{\Phi}}{2} \right)\\ = D_+ D_- \Phi + D_- D_+ \Phi + \alpha \, \sin\left(\frac{\Phi}{2} \right)\,.$$
Thus, if the right-hand side vanishes then the left-hand side vanishes and so the claim is verified. \par 
Writing the two fields out in components we have
\begin{align*}
& \widetilde{\Phi} =  \widetilde{X}+ \theta_- \, \widetilde{\psi}_+  + \theta_+ \, \widetilde{\psi}_- + \theta_- \theta_+ \, \widetilde{F}_{+-} + \cO(z)\,,\\
& \Phi =  X+ \theta_- \, \psi_+  + \theta_+ \, \psi_- + \theta_- \theta_+ \ F_{+-} + \cO(z)\,.
\end{align*}
The auto-B\"acklund transformation \eqref{eqn:BTc} tells us that both terms $\cO(z)$ are equal. Thus,
\begin{align*}
& \widetilde{\Phi} + \Phi = (\widetilde{X} + X) + \theta_- \, (\widetilde{\psi}_+  +\psi_+ ) + \theta_+ \, (\widetilde{\psi}_- + \psi_-) + \theta_- \theta_+ \, (\widetilde{F}_{+-} + F_{+-} )+ \cO(z)\,,\\
& \widetilde{\Phi} - \Phi = (\widetilde{X} - X) + \theta_- \, (\widetilde{\psi}_+  - \psi_+ ) + \theta_+ \, (\widetilde{\psi}_- - \psi_-) + \theta_- \theta_+ \, (\widetilde{F}_{+-} - F_{+-} )\,.
\end{align*}
The auto-B\"{a}cklund transformation \eqref{eqn:BTa}-\eqref{eqn:BTc} can then be written as
\begin{subequations}
\begin{align}
& D_-(\widetilde{\Phi} - \Phi)  =  2 a \, \lambda_+ \sin\left( \frac{\widetilde{\Phi} + \Phi}{4}\right)\, , \label{eqn:BT2a}\\
& D_+ (\widetilde{\Phi} + \Phi) = 2 a^{-1} \, \lambda_- \sin\left( \frac{\widetilde{\Phi} - \Phi}{4}\right)\,, \label{eqn:BT2b}
\end{align}
\end{subequations}
and so we require the $\cO(z)$ term to vanish, i.e., $Z_{-+} \widetilde{\Phi} = Z_{-+} \Phi =0$.  There is little loss in generality with this condition as, via mimicking standard superspace methods show that only the component fields independent of $z$ remain after performing the $\Z_2\times\Z_2$-graded Berezin integral (see \cite{Bruce:2020} for details). The equations \eqref{eqn:BT2a} and \eqref{eqn:BT2b} give a functional relationship between the $\Z_2 \times \Z_2$-superfields $\widetilde{\Phi}$ and $\Phi$. For small $a\in \R_*$ we can consider the Maclaurin--Taylor series 
\begin{equation}\label{eqn:MTseries}
\widetilde{\Phi} =  \sum_{n =0}^{\infty} a^n \, \widetilde{\Phi}_n(x, \theta, \lambda)\,.
\end{equation}
The auto-B\"{a}cklund transformations then take the form
\begin{subequations}
\begin{align}
&   \sum_{n=0}^\infty a^n \big(D_- \widetilde{\Phi}_n \big)-   D_- \Phi = 2 a \, \lambda_+ \sin \left( \frac{1}{4} \left( \sum_{n=0}^\infty a^n \widetilde{\Phi}_n + \Phi  \right) \right) \,,\label{eqn:BTExpa}\\
&   \sum_{n=0}^\infty a^n \big(D_+ \widetilde{\Phi}_n \big)+   D_+ \Phi = 2 a^{-1} \, \lambda_- \sin \left( \frac{1}{4} \left( \sum_{n=0}^\infty a^n \widetilde{\Phi}_n - \Phi  \right) \right) \,,\label{eqn:BTExpb}
\end{align}
\end{subequations}
Expanding and equating terms in order of $a$ allows us to deduce the relation between two $\Z_2\times \Z_2$-superfields. From \eqref{eqn:BTExpb} comparing the $\cO(a^{-1})$ terms (the left-hand side is clearly zero) we deduce that $\widetilde{\Phi}_0 = \Phi$. Comparing the $\cO(a^0)$ terms we deduce from \eqref{eqn:BTExpb} that $ 4 D_+ \Phi = \lambda_- \, \widetilde{\Phi}_1$. Then using the properties of the spinor-valued parameters we see that $\widetilde{\Phi}_1 =  - 4 (\lambda_+)^2 \lambda_- \, D_+ \Phi$. \par 
Moving on to $\cO(a)$ terms, from \eqref{eqn:BTExpa} and \eqref{eqn:BTExpb} we obtain the pair of equations
$$ D_-\widetilde{\Phi}_1 =  2 \lambda_+ \, \sin \left( \frac{\Phi}{2}\right)\,,\quad \textnormal{and} \quad  2 D_+ \widetilde{\Phi}_1 = \lambda_- \widetilde{\Phi}_2\,.$$
The first equation is consistent and contains no new information. Indeed, we observe that $D_- \widetilde{\Phi}_1 = - 4 (\lambda_+)^2 \lambda_- \, D_- D_+ \Phi$. By assumption we have no terms involving $z$ and so $D_-$ and $D_+$ commute and the $\Z_2 \times \Z_2$-graded sine-Gordon equation simplifies and so we have $D_- \widetilde{\Phi}_- =  2 (\lambda_+)^2 \lambda_- \alpha \sin(\Phi /2)= 2 \lambda_+ \sin (\Phi/2)$ using the algebraic condition on the spinor-valued parameters and their relation to $\alpha$. The second equation gives  $\widetilde{\Phi}_2 =  - 2 (\lambda_+)^2 \lambda_- D_+ \widetilde{\Phi}_1 = 8 (\lambda_+)^2 \, D_+ D_+ \Phi$. Then, for small $a \in \R_*$ we have 
\begin{equation}
\widetilde{\Phi} =  \Phi  -  4a (\lambda_+)^2\lambda_- \, D_+ \Phi + 8 a^2 (\lambda_+)^2 \, D_+ D_+ \Phi + \cO(a^3)\,.
\end{equation}
Continuing, we make a few observations.  We notice that, for $n \geq 1$, $\widetilde{\Phi}_n \propto D_+ \widetilde{\Phi}_{n-1}$ and thus $(\widetilde{\Phi}_n)^k =0$ for all $k\geq 2$. Then \eqref{eqn:BTExpb} tells us that
$$2 \,D_+ \widetilde{\Phi}_n = \lambda_- \, \widetilde{\Phi}_{n+1}\,,$$
which can be solved iteratively following our earlier calculations. Specifically, we obtain
\begin{equation}\label{eqn:PhinD}
\widetilde{\Phi}_n = (-1)^{n+1}\, 2^{n+1}\, (\lambda_+)^{2n}(\lambda_-)^n \, D^n_ +  \Phi \,,
\end{equation}
Then we can write 
\begin{equation}
\widetilde{\Phi} =  \Phi + \sum_{n=1}^\infty (-1)^{n+1}\, a^n \, 2^{n+1}\, (\lambda_+)^{2n}(\lambda_-)^n \, D^n_+ \Phi\,. 
\end{equation}
The system \eqref{eqn:BTExpa} is redundant and automatically satisfied. To see this, using the $\Z_2 \times\Z_2$-graded sine-Gordon equation together with the independence of   $\Phi$ on $z$ we obtain

\begin{align}\label{eqn:Dminis}\nonumber 
D_- \widetilde{\Phi}_n &= (-1)^{n-1} \, 2^n (\lambda_+)^{2n+1}(\lambda_-)^{n+1} \, D^{n-1}_+ \sin \big( \Phi / 2\big)\\ \nonumber 
& =(-1)^{n} \, 2\, \lambda_+ \,(\lambda_+)^{2n-1}(\lambda_-)^{n-1} \,\sum_{i=1} \sin^{(i)}\big(\Phi / 2 \big)\, D^{p_1}_+ \Phi \cdots D^{p_i}_+ \Phi\nonumber \\
&= 2 \, \lambda_+ \sum_{i = 1} \sin^{(i)}\big( \Phi / 2 \big) \widetilde{\Phi}_{p_1}\cdots \widetilde{\Phi}_{p_i}\,
 \end{align}
with $p_i > p_{i-1} \cdots > p_1$ and $p_i + \cdots + p_1 = n-1$. We recognise the final line of \eqref{eqn:Dminis} as being the $\cO(a^n)$ term of  \eqref{eqn:BTExpa}.
\section{Currents and conservation laws}
From the auto-B\"{a}cklund transformation \eqref{eqn:BT2a}-\eqref{eqn:BT2b}, we have the conserved (spinor) current
\begin{equation}
j_+ = a \lambda_+ \, \cos \left( \frac{\widetilde{\Phi} + \Phi}{4}\right)\,, \quad j_- = a^{-1} \lambda_- \, \cos \left( \frac{\widetilde{\Phi} - \Phi}{4}\right)\,. \label{eqn:Current}
\end{equation}
Note that $j_+$ and $j_-$ are of $\Z_2 \times \Z_2$-degree $(0,1)$ and $(1,0)$, respectively. Under Lorentz transformations we have 
$$j_+ \mapsto \rme^{\half \beta}\, j_+\, , \quad j_- \mapsto \rme^{- \half \beta} \, j_-\,.$$
We show that we do indeed have a conserved current in this way via direct calculation:
\begin{align*}
 D_- j_-  + D_+ j_+  &=  - \frac{a^{-1}}{4}\lambda_- \, \sin \left( \frac{\widetilde{\Phi} - \Phi}{4}\right)\, (D_- \widetilde{\Phi} - D_- \Phi) - \frac{a}{4}\lambda_+ \, \sin \left( \frac{\widetilde{\Phi} + \Phi}{4}\right)\, (D_- \widetilde{\Phi} + D_- \Phi)\\
&= - \frac{\lambda_- \lambda_+ }{2}\, \sin \left( \frac{\widetilde{\Phi} - \Phi}{4}\right)\sin \left( \frac{\widetilde{\Phi} + \Phi}{4}\right)
 - \frac{\lambda_+ \lambda_-}{2} \, \sin \left( \frac{\widetilde{\Phi} + \Phi}{4}\right)\sin \left( \frac{\widetilde{\Phi} - \Phi}{4}\right)\\
& =\frac{\alpha}{2} \left( \sin \left( \frac{\widetilde{\Phi} - \Phi}{4}\right)\sin \left( \frac{\widetilde{\Phi} + \Phi}{4}\right)
  - \sin\left( \frac{\widetilde{\Phi} - \Phi}{4}\right)\sin \left( \frac{\widetilde{\Phi} + \Phi}{4}\right)\right)\\
&=0.
\end{align*}
From the current \eqref{eqn:Current} we have the conservation law
\begin{equation}\label{eqn:ConLaws}
\lambda_+ D_-\left( \cos \left(\frac{\widetilde{\Phi} + \Phi}{4} \right)\right)  = - a^{-2}\, \lambda_- D_+ \left( \cos \left(\frac{\widetilde{\Phi} - \Phi}{4} \right)\right)\,.
\end{equation}
We can then consider the expansion for small $a$ (see \eqref{eqn:MTseries}). 
In doing so we have a countable infinite set of conservation rules defined order-by-order in $a$. We observe that the right-hand-side of \eqref{eqn:ConLaws} vanishes as $(D_+ \Phi)^k= 0$ if $k \geq 2$. Then using \eqref{eqn:PhinD} we deduce that
\begin{equation}\label{eqn:ConLaws1}
D_- \left(\cos\big(\Phi / 2 \big) \right) =0\, , \quad D_- \left( \sin\big(\Phi / 2  \big) \, (D_+)^k \Phi \right) =0\,, \quad k \geq 1\,,
\end{equation}
and thus we have an infinite  set of conservation laws. Repeating the analysis of this paper with the other natural choice of auto-B\"{a}cklund transformation (as the $\Z_2\times \Z_2$-graded sine-Gordon equation is symmetric under exchange of $D_-$ and $D_+$)
 \begin{subequations}
\begin{align}
& D_+\widetilde{\Phi} =  D_+ \Phi + 2 a \, \eta_+ \sin\left( \frac{\widetilde{\Phi} + \Phi}{4}\right)\,, \label{eqn:BT3a}\\
&  D_-\widetilde{\Phi} = - D_- \Phi + 2 a^{-1} \, \eta_- \sin\left( \frac{\widetilde{\Phi} - \Phi}{4}\right)\,, \label{eqn:BT3b}\\
& Z_{-+ }\widetilde{\Phi} =  Z_{-+}\Phi\,, \label{eqn:BT3c}
\end{align}
\end{subequations}
we obtain another countably infinite set of conservation laws
\begin{equation}\label{eqn:ConLaws2}
D_+ \left(\cos\big(\Phi / 2 \big) \right) =0\, , \quad D_+ \left(  \sin\big(\Phi / 2  \big)\, (D_-)^k \Phi \right) =0\,, \quad k \geq 1\,.
\end{equation}
\section{Concluding remarks}
Via construction of  auto-B\"{a}cklund transformations for the $\Z_2 \times \Z_2$-graded sine-Gordon equation we built a pair of infinite sets of conservation laws (see \eqref{eqn:ConLaws1} and \eqref{eqn:ConLaws2}). Thus, the $\Z_2 \times \Z_2$-graded sine-Gordon equation is integrable. The auto-B\"{a}cklund transformations involve  pairs of spinor-valued parameters  that mutually anticommute that are related to the $(1,1)$-degree parameter. The presence of spinor-values parameters was expected as we are dealing with a generalisation of supersymmetry. The extra algebraic conditions are forced upon us as the $\Z_2 \times \Z_2$-graded sine-Gordon equation has a Clifford algebra-valued coupling constant of $\Z_2 \times \Z_2$- degree $(1,1)$. Such coupling constants are forced on us if we want an interacting theory of degree zero $\Z_2\times \Z_2$-superfields, this properly discussed in \cite{Bruce:2020}. \par 
It would be interesting to check if a similar analysis can be performed on $\Z_2 \times \Z_2$-graded versions of  (supersymmetric) sinh-Gordon, Liouville  or more generally Toda models. It is expected that appropriate auto-B\"{a}cklund transformations can be found that involve spinor-valued parameters. \par

We stress that the analysis here is purely classical and it remains to be seen what survives quantisation and if any subtleties of quantisation with $\Z_2 \times \Z_2$-graded fields complicate the picture. That said, supersymmetry usually improves the quantum properties of a theory compared to the related pure bosonic theory. One conjectures the same is true of $\Z_2 \times \Z_2$-supersymmetry.

\section*{Acknowledgements} 
The author thanks  Steven Duplij and Francesco Toppan for their support and comments on earlier drafts of this paper.  


\begin{thebibliography}{10}
\begin{small}
\bibitem{Aizawa:2020a}
N.~Aizawa,  K.~Amakawa \& S.~Doi,
$\mathcal N$-extension of double-graded supersymmetricand superconformal quantum mechanics, \href{https://doi.org/10.1088/1751-8121/ab661c}{\emph{J. Phys. A: Math. Theor.}} \textbf{53} (2020), 065205.

\bibitem{Aizawa:2020}
N.~Aizawa, Z.~Kuznetsova \& F.~Toppan,
$\Z_2 \times \Z_2$-graded mechanics: The classical theory, \href{https://doi.org/10.1140/epjc/s10052-020-8242-x}{\emph{Eur. J. Phys. C}} \textbf{80} (2020), 668. 

\bibitem{Aizawa:2021}
N.~Aizawa, Z.~Kuznetsova \& F.~Toppan,
$\Z_2 \times \Z_2$-graded mechanics: The quantization,
\href{https://doi.org/10.1016/j.nuclphysb.2021.115426}{\emph{Nuclear Physics B}} \textbf{967} (2021), 115426.

\bibitem{Barone:1971}
A.~Barone, F.~Esposito, C.J.~Magee \& A.C.~Scott, 
Theory and applications of the sine-gordon equation,
\href{https://doi.org/10.1007/BF02820622}{\emph{La Rivista del Nuovo Cimento (1971-1977)}} volume 1 (1971), pages 227--267.

\bibitem{Bertrand:2017}
S.~Bertrand, 
On integrability aspects of the supersymmetric sine-Gordon equation,
\href{https://doi.org/10.1088/1751-8121/aa6324}{\emph{J. Phys. A}} \textbf{50} (2017), no. 16, 165202, 14 pp.


\bibitem{Bruce:2019a}
A.J.~Bruce, On a $\Z_2^n$-graded version of supersymmetry, \href{https://doi.org/10.3390/sym11010116}{\emph{Symmetry}} 2019, \textbf{11}, 116. 

\bibitem{Bruce:2020}
A.J.~Bruce,
$\Z_2 \times \Z_2$-graded supersymmetry: 2-d sigma models,
\href{https://doi.org/10.1088/1751-8121/abb47f}{\emph{J. Phys. A}} \textbf{53} (2020), no. 45, 455201, 25 pp.

\bibitem{Bruce:2020a}
A.J.~Bruce \& S.~Duplij,
Double-graded supersymmetric quantum mechanics, \href{ https://doi.org/10.1063/1.5118302}{\emph{J. Math.Phys.}} \textbf{61} (2020), 063503.

\bibitem{Chainchain:1978}
M.~Chaichian \& P.P.~Kulish,
On the method of inverse scattering problem and Bäcklund transformations for supersymmetric equations,
\href{https://doi.org/10.1016/0370-2693(78)90473-2}{\emph{Phys. Lett. B}}, \textbf{78} (1978), Issue 4, 413--416.

\bibitem{Covolo:2016}
T.~Covolo, J.~Grabowski \& N.~Poncin,
The category of $\Z_2^n$-supermanifolds,
\href{https://doi.org/10.1063/1.4955416}{\emph{J. Math. Phys.}} \textbf{57} (2016), no. 7, 073503, 16 pp. 

\bibitem{DiVecchia:1977}
P.~Di~Vecchia \& S.~Ferrara,
Classical solutions in two-dimensional supersymmetric field theories,
\href{https://doi.org/10.1016/0550-3213(77)90394-7}{\emph{Nuclear Physics B}} \textbf{130} (1977), Issue 1, 93--104.

\bibitem{Duplij:2004}
 S.~Duplij, W.~Siegel \& J.~Bagger (editors),
\href{https://doi.org/10.1007/1-4020-4522-0}{Concise encyclopedia of supersymmetry and noncommutative structures in mathematics and physics}, \emph{Kluwer Academic Publishers, Dordrecht}, 2004. iv+561 pp. ISBN: 1-4020-1338-8 

\bibitem{Evans:1992}
J.~Evans \& T.~ Hollowood,
Integrable $N = 2$ supersymmetric field theories,
\href{https://doi.org/10.1016/0370-2693(92)91486-S}{\emph{Phys. Lett. B}}
\textbf{293} (1992), Issues 1--2, 100--110.

\bibitem{Ferrara:1978}
S. Ferrara, L. Girardello \& S. Sciuto,
An infinite set of conservation laws of the supersymmetric sine-gordon theory \href{https://doi.org/10.1016/0370-2693(78)90793-1}{\emph{Phys. Lett. B}} \textbf{76} (1978), Issue 3, 303--306.

\bibitem{Grigoriev:2008}
M.~Grigoriev \& A.A.~Tseytlin,
Pohlmeyer reduction of $AdS_5 \times S_5$ superstring sigma model,
\href{https://doi.org/10.1016/j.nuclphysb.2008.01.006}{\emph{Nuclear Physics B}} \textbf{800} (2008), Issue 3, 450--501,

\bibitem{Ivanov:1993}
E.~Ivanov \& F.~Toppan,
N = 2 superconformal affine Liouville theory,
\href{https://doi.org/10.1016/0370-2693(93)90936-C}{\emph{Phys. Lett. B}}
\textbf{309} (1993), Issues 3--4, 289--296.


\bibitem{Kranner:1991}
G.~Kranner \& W.~Kummer,
Universal conditions of finiteness for general quantum field theories,
\href{https://doi.org/10.1016/0370-2693(91)90138-G}{\emph{Phys. Lett. B}} \textbf{259} (1991), Issues 1--2, 84--90.

\bibitem{Kuznetsova:2008}
Z.~Kuznetsova, M.~Rojas, \& F.~Toppan, On supergroups with odd Clifford parameters and supersymmetry with modified Leibniz rule, \href{https://doi.org/10.1142/S0217751X08038159}{\emph{Internat. J. Modern Phys.}} \textbf{A23} (2008), no. 2, 309--326.


\bibitem{Liu:1998}
Q.P.~Liu \& M.~Ma\~nas,
Pfaffian solutions for the Manin-Radul-Mathieu SUSY KdV and SUSY sine-Gordon equations,
\href{https://doi.org/10.1016/S0370-2693(98)00852-1}{\emph{Phys. Lett. B}}
\textbf{436} (1998), Issues 3--4,  306--310.

\bibitem{Poncin:2016}
N.~Poncin,
Towards integration on colored supermanifolds, in \href{https://doi.org/10.4064/bc110-0-14}{\emph{Geometry of jets and fields}}, 201--217, Banach Center Publ., \textbf{110}, \emph{Polish Acad. Sci. Inst. Math., Warsaw}, 2016.


\bibitem{Rittenberg:1978}
V.~Rittenberg \& D. ~Wyler, Generalized Superalgebras,
 \href{https://doi.org/10.1016/0550-3213(78)90186-4}{\emph{Nucl. Phys. B}} \textbf{139} (1978), 189--202.

\bibitem{Scheunert:1979}
 M.~Scheunert, Generalized Lie algebras, 
 \href{https://doi.org/10.1063/1.524113}{\emph{J. Math. Phys.}} \textbf{20} (1979), 712.

\bibitem{Tolstoy:2014}
V.N.~Tolstoy, Once more on parastatistics, \href{ https://doi.org/10.1134/S1547477114070449}{\emph{Phys. Part. Nuclei Lett.}} \textbf{11} (2014), 933--937.

\bibitem{Toppan:2021}
F.~Toppan, $\Z_2^2$-graded parastatistics in multiparticle quantum Hamiltonians, \href{https://doi.org/10.1088/1751-8121/abe2f2}{\emph{J. Phys. A: Math. Theor.}} \textbf{54} (2021), 115203.


\end{small}
\end{thebibliography}
\end{document}